\documentclass[twocolumn,showpacs]{revtex4-1}
\usepackage{graphicx}
\usepackage{amsmath}
\usepackage{booktabs}
\begin{document}
\title{Strong ties promote the epidemic prevalence in susceptible-infected-susceptible spreading dynamics}

\author{Ai-Xiang Cui}
\author{Zimo Yang}
\author{Tao Zhou}
\email{zhutou@ustc.edu}

\affiliation{
Web Sciences Center, University of Electronic Science and Technology of China, Chengdu 611731, People's Republic of China
}

\begin{abstract}
Understanding spreading dynamics will benefit society as a whole in better preventing and controlling diseases, as well as facilitating the socially responsible information while depressing destructive rumors. In network-based spreading dynamics, edges with different weights may play far different roles: a friend from afar usually brings novel stories, and an intimate relationship is highly risky for a flu epidemic. In this article, we propose a weighted susceptible-infected-susceptible model on complex networks, where the weight of an edge is defined by the topological proximity of the two associated nodes. Each infected individual is allowed to select limited number of neighbors to contact, and a tunable parameter is introduced to control the preference to contact through high-weight or low-weight edges. Experimental results on six real networks show that the epidemic prevalence can be largely promoted when strong ties are favored in the spreading process. By comparing with two statistical null models respectively with randomized topology and randomly redistributed weights, we show that the distribution pattern of weights, rather than the topology, mainly contributes to the experimental observations. Further analysis suggests that the weight-weight correlation strongly affects the results: high-weight edges are more significant in keeping high epidemic prevalence when the weight-weight correlation is present.

\end{abstract}


\date{\today}

\maketitle

\section{Introduction}

Early before the classification of social ties proposed, in 1954, the Russian mathematical psychologist Anatol Rapoport \cite{Rapoport1954} has been aware of ``well-known fact that the likely contacts of two individuals who are closely acquainted tend to be more overlapping than those of two arbitrarily selected individuals". This argument became one cornerstone of social network theory. In 1973, ties in social networks, generally, come in two varieties: strong and weak, which has been first proposed by the American sociologist Mark Granovetter \cite{Granovetter1973}. Different relationships can be measured in the currency of tie strength. According to the closeness, connections to close friends have been said to be ``strong'' ties, while those to acquaintances have been called ``weak'' ties \cite{Granovetter1973,Erickson1978,Lin1986,Granovetter1995}.

Strong ties connect with the people you really trust, people whose social circles tightly overlap with your own. Often, they are also the people most like you. Weak ties, conversely, connect with merely acquaintances and often provide access to novel information. Tie strength usually plays an vital role in many real networks and is crucial to understand dynamical processes on the networks \cite{Shi2007,Barrat2008}. Weak ties display an important bridging function \cite{Zhao2010,Cheng2010}, while strong ties are more likely to activate the flow of referral information and more influential than weak ties \cite{Brown1987,Steffes2009}. In addition, weak ties could play a more significant role than strong ties to keep the stability \cite{Csermely2006}, maintain the connectivity \cite{Onnela2007} and uncover the missing information \cite{Lu2010,Lu2011}, while strong ties can be better utilized to enhance the human resource flexibility \cite{Zolin2011}, to provide accurate recommendation \cite{Medo2009,Ye2010,Lu2012}, and so on.

In despite of the qualitative distinction between strong and weak ties, tie strength could be quantitatively described by edge weight---the edges with high weights are considered to be strong. In a number of social networks, edges are often associated with weights that differentiate them in terms of their strength, intensity, capacity or the frequency of recent contacts \cite{Granovetter1973,Barrat2004}. For non-social networks, weights often refer to the functions performed by edges, e.g., the amount of traffic flowing along connections in world-wide airport networks \cite{Barrat2004}, the number of joint papers of two coauthors in scientific collaboration networks \cite{Barrat2004}, the number of synapses and gap junctions in neural networks \cite{Rojas1996}, the carbon flow between species in food webs \cite{Nordlund2007}.

Weight plays a significant role in disparate network-based dynamics, such as transportation \cite{Wang2005,Yan2006}, synchronization \cite{Chavez2005,Zhou2006,Arenas2008}, percolation \cite{Wu2006,Li2007}, and so on. In this article, we concentrate on the effects of weights on epidemic spreading. Yan \textit{et al.} \cite{Yan2005} investigated the epidemic spreading in weighted scale-free networks and the simulation results indicated that the more homogeneous weight distribution of the network is, the more quickly epidemic spreads on it. This finding was further demonstrated by an edge-based mean-field solution \cite{Yang2012}. Chu \textit{et al.} \cite{Chu2011} showed that weight distribution has strong impacts on both epidemic threshold and prevalence. Karsai \emph{et al.} \cite{Karsai2006} considered the contact process in weighted scale-free networks, in which the weight of an edge connecting two high-degree nodes is relatively small. Yang \emph{et al.} \cite{Yang2008} further proved that in the contact process, the epidemic prevalence can be maximized by setting the edge weight inversely proportional to the degree of the receiving node. Baronchelli and Pastor-Satorras \cite{Baronchelli2010} considered the diffusive dynamics on weighted networks and shed light on the validity of mean-field theory on weighted networks. Rattana \textit{et al.} \cite{Rattana2013} proposed a pairwise-type approximation for epidemic dynamics on weighted networks, showing a more accurate solution than traditional methods.

In this article, we propose a weighted susceptible-infected-susceptible (SIS) epidemic spreading model, with a tunable parameter controlling the preference of spreading: whether or not an infected individual prefers to contact others through edges with high weights. Experimental results on six real networks show that the preferential contacts through strong ties could largely improve the epidemic prevalence. We compare such results with two statistical null models, where the topological structure and weight distribution are randomized respectively. The aforementioned strong ties effects is qualitatively the same under randomized topology while is vanished if the weights are randomly redistributed, indicating that the distribution pattern of weights mainly contributes to the experimental observations. Further analysis suggests that the weight-weight correlation strongly affects the results, with the optimal value of the controlling parameter monotonously depending on the correlation strength.

This article is organized as follows. In Section II, we present the details of the SIS model. The data description, together with simple statistics, are shown in Section III. Section IV reports the experimental results and Section V provides theoretical insights by comparing the experimental results with statistical null models. The main conclusions are summarized in Section VI.

\section{Model}

The SIS model \cite{Anderson1992,Hethcote2000} is suitable to describe the cases when individuals cannot acquire immunity after recovering from the disease, such as influenza, pulmonary tuberculosis and gonorrhea. With disease of this kind individuals that are cured usually catch again. In the networked SIS model, nodes are in two discrete states, ``susceptible'' or ``infected''. Each infected node will contact all its neighbors once at each time step, and therefore the infectivity of each node is proportional to its degree. In the real world, individuals may be only able to contact limited population within one time step \cite{Zhou2006b}. For example, salesman in network marketing processes will not make referrals to all his acquaintances due to the limited money and time \cite{Kim2006}. In sexual contact networks, although a few individuals have hundreds of sexual partners, their sexual activities are not far beyond a normal level due to the physiological limitations \cite{Liljeros2003,Schneeberger2004}. Therefore, in the present model we assume every individual has the same infectivity \cite{Yang2007,Zhou2006b}. Without the lose of generality, at each time step, each infected node will select one of its neighbors to contact. If the selected neighbor has been infected already, nothing happens, otherwise it will be infected with probability $\alpha$. Meanwhile, each infected node will become susceptible in the next time step with probability $\beta$. Since each infected individual only selects one neighbor at each time step, the disease can spread out only when $\alpha>\beta$ \cite{Yang2007}. In the following study, we fix $\alpha=0.4$ and $\beta=0.1$, and we have already checked that the specific choices of $\alpha$ and $\beta$ will not change the qualitative results reported in this article.

\begin{table}
\caption{\label{basic-characteristic}
Basic structural features of the largest connected components of the studied six real networks. $N$ and $E$ are the number of the nodes and edges, $k_{max}$ and $\langle k \rangle$ are the maximum degree and average degree over all nodes. $\langle d \rangle$ is the average shortest path length, $C$ and $r_D$ are the clustering coefficient \cite{Watts1998} and degree-degree correlation coefficient \cite{Newman2002} respectively.
}
\begin{tabular}{cccccccc}
\hline
Networks & $N$ & $E$ & $k_{max}$ & $\langle k \rangle$ & $\langle d \rangle$ & $C$ & $r_D$ \\
\hline
FSN & 1893 & 13835 & 255 & 14.6 & 3.06 & 0.11 & -0.188 \\
EEN & 33696 & 180811 & 1383 & 10.7 & 4.03 & 0.509 & -0.116 \\
SSN & 74444 & 382456 & 2517 & 10.3 & 4.21 & 0.06 & -0.067 \\
ESN & 75868 & 405729 & 3044 & 10.7 & 4.31 & 0.138 & -0.041\\
GPN & 8842 & 31837 & 88 & 7.2 & 4.60 & 0.007 & 0.015\\
OAS & 10670 & 22002 & 2312 & 4.1 & 3.64 & 0.297 & -0.186\\
\hline
\end{tabular}
\end{table}

\begin{figure}
\includegraphics[width=8cm]{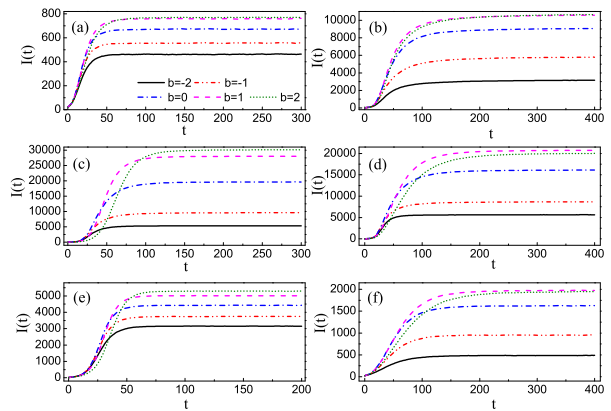}
\caption{(Color online) Spreading processes for different values of the parameter $b$. These six plots show how the number of infected individuals, $I(t)$, changes with time on the six real networks: (a) Facebook-like social network, (b) Enron email network, (c) Slashdot social network, (d) Epinions social network, (e) Gnutells peer-to-peer network, (f) Oregon autonomous systems. Results are obtained by averaging over 1000 independent realizations.}
\end{figure}

The probability that an infected node $i$ will select its neighbor $j$ is
\begin{equation}
p_{ij}=\frac{s^{b}_{ij}}{\sum_{l\in{\Gamma_{i}}}s^{b}_{il}},
\end{equation}
where $\Gamma_{i}$ is the set of neighbors of node $i$, $s_{ij}$ denotes the tie strength between $i$ and $j$, and $b$ is a tunable parameter. If $b=0$, the infected node randomly selects a neighbor to contact, equivalent to an unweighted SIS model \cite{Yang2007}. If $b>0$, strong ties are favored to constitute the paths of spreading, while if $b<0$, weak ties are favored.

In different contexts, the strength of a tie may have different definitions and measures \cite{Marsden1984,Lu2011}, which may depend on external information to network topology. For general networks, one may be not aware of external information and thus it is meaningful to give a natural definition solely based on network topology. According to Rapoport's theory \cite{Rapoport1954} and other supportive observations \cite{Kossinets2006,Liben2007,Zhou2009} and models \cite{Holme2002,Cui2011,Cannistraci2013}, we define the tie strength $s_{ij}$ in spite of $i$ and $j$'s common neighbors, as follows:
\begin{equation}\label{sij}
s_{ij}=\frac{n_{ij}+\delta}{\sqrt{k_{i}k_{j}}},
\end{equation}
where $n_{ij}$ is the number of common neighbors of $i$ and $j$, $k_i$ is the degree of $i$, and $\delta$ is a constant that gives chance to the tie connecting two nodes without common neighbors. For simplicity, we set $\delta=1$. The dynamical process starts with randomly selecting a certain number of infected nodes, we set it as 20. This initial number has no effect on the stable state.

\begin{figure}
\includegraphics[width=8cm]{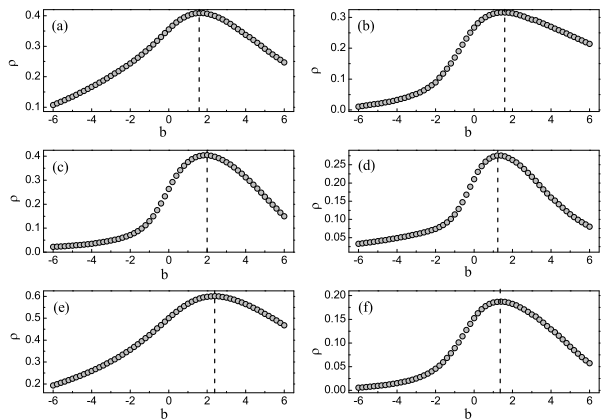}
\caption{The epidemic prevalence $\rho$ as a function of the parameter $b$ based on the six real networks. These six plots correspond to (a) Facebook-like social network, (b) Enron email network, (c) Slashdot social network, (d) Epinions social network, (e) Gnutells peer-to-peer network, and (f) Oregon autonomous systems. Results are obtained by averaging over 1000 independent realizations.}
\end{figure}

\section{Data}

To see the impacts of tie strength on the spreading dynamics, experiments are carried out on six real networks: (1) Facebook-like Social Network (FSN) \cite{Opsahl2009}: it originates from an online community for students at University of California. The data set includes the users that have sent or received at least one message, and an undirected edge is set between $i$ and $j$ if $i$ has sent (received) an online message to (from) $j$. (2) Enron Email Network (EEN) \cite{Leskovec2009,Klimt2004}: it covers about half million emails. Nodes of this network are email addresses and if an address $i$ sent at least one email to address $j$, an undirected edge is established between $i$ and $j$. (3) Slashdot Social Network (SSN) \cite{Leskovec2010}: nodes in this network are the users in Slashdot, which is a technology-related news website, and edges represent friendships between users. (4) Epinions Social Network (ESN) \cite{Richardson2003}: this is a who-trust-whom online social network. Nodes are the members of the general consumer review site \emph{Epinions.com}, and edges represent the trust relationships between two members. (5) Gnutella Peer-to-peer Network (GPN) \cite{Ripeanu2002}: Nodes represent hosts and edges stand for connections between the hosts. (6) Oregon Autonomous Systems (OAS) \cite{Leskovec2005}: this is an AS-level Internet topology graph obtained by the \emph{Route Views Project}. In order to guarantee the connectivity, we use the largest connected components of these networks.
Table \ref{basic-characteristic} presents the basic statistics of the largest connected components of the six networks.

\begin{figure}
\includegraphics[width=8cm]{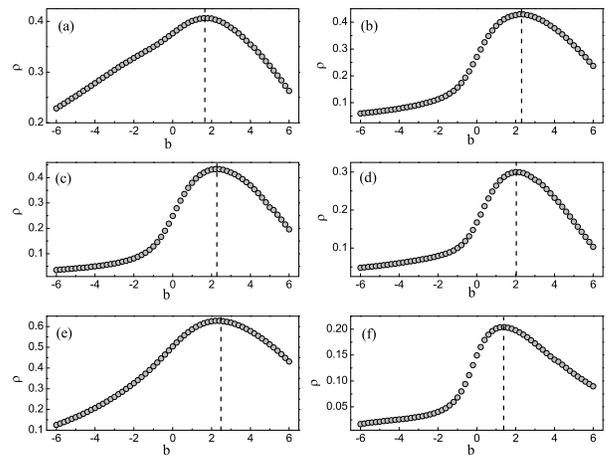}
\caption{The epidemic prevalence $\rho$ as a function of the parameter $b$ based on the configuration model corresponding to (a) Facebook-like social network, (b) Enron email network, (c) Slashdot social network, (d) Epinions social network, (e) Gnutells peer-to-peer network, and (f) Oregon autonomous systems. Results are obtained by averaging over 1000 independent realizations.}
\end{figure}

\section{Experimental Results}

Figure 1 gives us an intuition of the spreading processes following the proposed model, showing how the number of infected nodes, $I(t)$, changing with time, $t$. Clearly, the epidemic dynamics is considerably affected by the parameter $b$, in a complicated and non-monotonous way. Define the epidemic prevalence as the fraction of infected nodes in the stable state
\begin{equation}
\rho=\frac{I(\infty)}{N}.
\end{equation}
Figure 2 displays how the epidemic prevalence $\rho$ changes with $b$ for the six real networks. For each case, there is a well-defined peak for $\rho(b)$. For convenience, we call the value of $b$ corresponding to the largest prevalence $\rho^*$ the optimal value of $b$, denoted by $b^*$, .

\begin{figure}
\includegraphics[width=8cm]{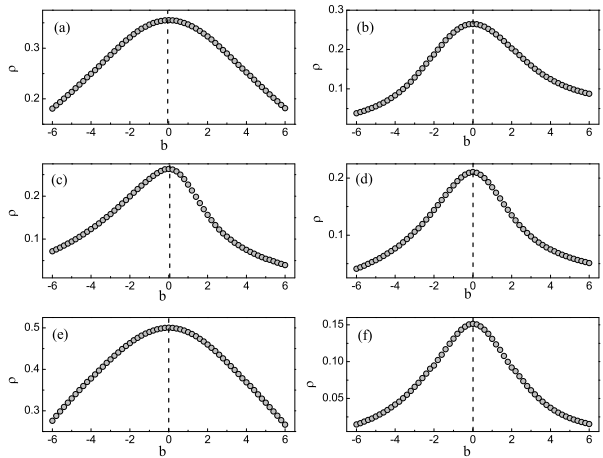}
\caption{The epidemic prevalence $\rho$ as a function of the parameter $b$ based on the null model with randomly redistributed weights that correspond to (a) Facebook-like social network, (b) Enron email network, (c) Slashdot social network, (d) Epinions social network, (e) Gnutells peer-to-peer network, and (f) Oregon autonomous systems. Results are obtained by averaging over 1000 independent realizations.}
\end{figure}

As shown in figure 2, for all the six real networks, the optimal value $b^*>0$. Comparing with the unweighted case $b=0$, the epidemic prevalence for each real network is promoted remarkably. Sometimes, it almost gets doubled. In a word, the epidemic prevalence can be largely promoted when strong ties are favored in the spreading process.

\section{Analysis}

To validate the significance of our findings and dig out underlying contributors for the nontrivial peaks in $\rho(b)$ curves, we further compare our results with two statistical null models. The first one is called configuration model \cite{Catanzaro2005}. Given a real network with degree sequence $\{k_1,k_2,\cdots,k_N\}$, where $k_i$ is the degree of the $i$th node. In the configuration model, the $i$th node is assigned $k_{i}$ stubs and the network is constructed by randomly choosing stubs and connecting them to form edges, avoiding multiple connections and self-connections. Detailed rules can be found in Ref. \cite{Catanzaro2005}. Under the configuration model, the topology, except for the degree heterogeneity, is randomized, while the correlation between local structure and weight is held, since an edge's weight is still determined by the common neighborhood and degree of its two endpoints, by Eq. (2). As shown in figure 3, in despite of the slightly different values of $b^*$ in original and randomized networks, the $\rho(b)$ curves in configuration models are qualitatively the same to original ones.

\begin{figure}
\includegraphics[width=8cm]{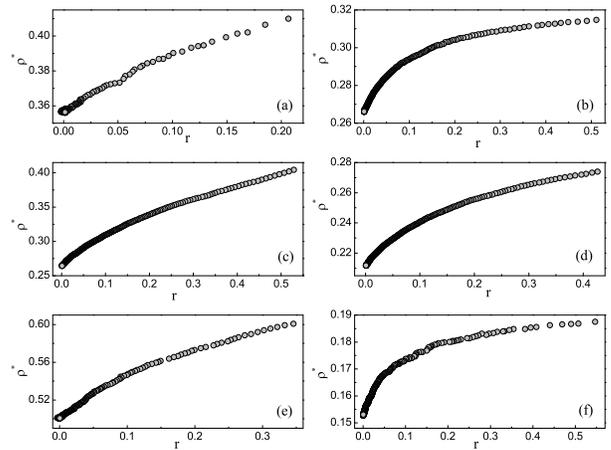}
\caption{The maximal epidemic prevalence $\rho^*$, corresponding to the optimal value $b^*$, as a function of the weight-weight correlation $r$. These six plots are obtained during the weight exchanging processes of the six real networks: (a) Facebook-like social network, (b) Enron email network, (c) Slashdot social network, (d) Epinions social network, (e) Gnutells peer-to-peer network, and (f) Oregon autonomous systems. For a given $r$, the corresponding $\rho^*$ is an average over 100 independent realizations.}
\end{figure}

In the second null model, for a given real network, we keep its topology unchanged and calculate edge weights according to Eq. (2), then we exchange the weights of two randomly selected edges \cite{Li2005,Li2007b}. After sufficient number of exchanges (we set it as $10E$ in this article), the weights are randomly redistributed. Under this null model, the topology is held while the correlation between local structure and edge weight is vanished. As shown in figure 4, after the redistribution of weights, the optimal value is exactly $b^*=0$ for all six cases. At this point, the weighted SIS model degenerates to the unweighted one, or it is equivalent to the most homogeneous case with all weights are the same. Therefore, the results are in accordance with the known conclusion \cite{Yang2012} that in the absence of correlation between structure and weight, the most homogeneous weight distribution leads to the widest epidemic spreading.

In a word, by comparing with these two statistical null models, we conclude that the distribution pattern of weights, or say the correlation between local structure and weight, rather than the topology itself, mainly contributes to the experimental observations.

The correlation of weights of two adjacent edges continuously changes in the randomizing process of the second null model. Here we use the well-known Pearson correlation coefficient $r$ \cite{Rodgers1988}, similar to the degree-degree correlation coefficient \cite{Newman2002}, to quantify the weight-weight correlation. Denote by $\Gamma=\{(e_{A_1},e_{B_1}),(e_{A_2},e_{B_2}),\cdots,(e_{A_M},e_{B_M})\}$ the set of all edge pairs sharing a common endpoint (i.e., two adjacent edges), where $M=\sum^N_{i=1}k_i(k_i-1)$ is the number of these edge pairs. Then the weight-weight correlation coefficient is defined as
\begin{equation}
r=\frac{\sum_{(e_A,e_B)\in \Gamma} (s_A-\overline{s})(s_B-\overline{s})}{\sqrt{\sum_{(e_A,e_B)\in \Gamma} (s_A-\overline{s})^2} \sqrt{\sum_{(e_A,e_B)\in \Gamma} (s_B-\overline{s})^2}},
\end{equation}
where the sum runs over all $M$ pairs and $s_A$ and $s_B$ denote the weights of edges $e_A$ and $e_B$, respectively. Note that, if $(e_A,e_B)\in \Gamma$, then $(e_B,e_A)\in \Gamma$ too. Therefore $\overline{s}=\frac{1}{M}\sum^M_{i=1}s_{A_i}=\frac{1}{M}\sum^M_{i=1}s_{B_i}$. For some $i$ and $j$, it is possible that $e_{A_i}\equiv e_{A_j}$, and for an edge $e$, if its two associated nodes are respectively of degree $k^1_e$ and $k^2_e$, the weight $w_e$ has been counted $(k^1_e-1)+(k^2_e-1)$ times. Clearly, the value of $r$ lies in the range $[-1,1]$: $r>0$ indicates a positive correlation (i.e., high-weight edges tend to be adjacent to other high-weight edges), $r<0$ indicates a negative correlation (i.e., high-weight edges tend to be adjacent to low-weight edges), and $r=0$ is for the case of no correlation.

\begin{figure}
\includegraphics[width=8cm]{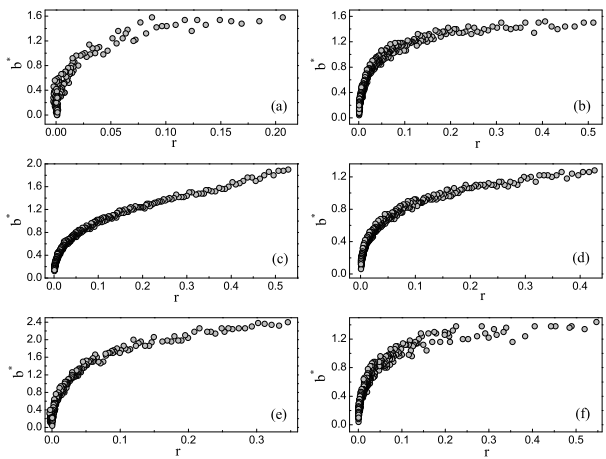}
\caption{The optimal value $b^*$, as a function of the weight-weight correlation $r$. These six plots are obtained during the weight exchanging processes of the six real networks: (a) Facebook-like social network, (b) Enron email network, (c) Slashdot social network, (d) Epinions social network, (e) Gnutells peer-to-peer network, and (f) Oregon autonomous systems. For a given $r$, the corresponding $b^*$ is an average over 100 independent realizations.}
\end{figure}

For all the six real networks, the weight-weight correlation coefficients are all considerably larger than zero, and during the randomizing process of the second null model, $r$ almost monotonously decays to zero. As shown in figure 5 and figure 6, both the maximal epidemic prevalence $\rho^*$ (corresponding to $b^*$) and the optimal value $b^*$ monotonously change with $r$. In particular, when $r=0$, the optimal value of $b$ is also equal to zero, indicating that the nontrivial peak comes from the local correlation of weight distribution. Additionally, in the real networks, compared with the null model, we can enlarge the prevalence by subtly adjusting the system. If the system is adjustable, this is indeed a good news when considering the spreading of valuable information and innovation, while we can assign $b$ a very large absolute value to against the infectious disease. Real systems never work in such a perfectly mathematical way, but our analysis provides some theoretical insights that may benefit the design of interventions of real spreading dynamics.

\section{Conclusion and Discussion}

Experimental results on six real networks demonstrated that the preferential contacts through high-weight edges can promote the epidemic prevalence in the stable state. Further analysis showed that this phenomenon results from the correlation between structure and weight, rather than the topological structure itself. Specifically speaking, it is contributed by the local correlation of weight distribution pattern. Although the non-zero weight-weight correlation requires a heterogeneous distribution of weights, the heterogeneity of weight distribution alone will not lead to the aforementioned non-trivial phenomenon.

Although the spreading processes in real systems are far different from the present ideal model, this study gives some inspirations on how to prevent infectious diseases and facilitate socially responsible information. However, even if in an ideal system, the design of interventions is very challenging since the different goals may conflict to each other. In the present model, the preference towards strong ties could enlarge the prevalence while it may slow down the spreading process (details will be reported elsewhere). Analogous examples are numerous: bypassing main intersections and arterial roads could enhance the network throughput while increase the delivering time especially in uncongested states \cite{Yan2006}, and removing directed loops could enhance the synchronizability while slow down the synchronizing process \cite{Zhou2010}.

Different nodes (e.g., hub nodes and peripheral nodes) and different edges (e.g., strong ties and weak ties) are supposed to play different roles in the evolution and functions of the network. By introducing the structure-based weights of edges, we can to some extent distinguish the roles of different edges in the spreading dynamics. We hope this work could contribute to the long journey towards fully understanding the relations between structural features and functional performances.

\section*{Acknowledgments}
We acknowledge Pak-Ming Hui for valuable discussions. This work is partially supported by the National Natural Science Foundation of China under Grant No. 11222543 and the Special Project of Sichuan Youth Science and Technology Innovation Research Team under Grant No. 2013TD0006. A.-X.C. acknowledges the Fundamental Research Funds for the Central Universities under Grant No. ZYGX2012YB027 and T.Z. acknowledges the Program for New Century Excellent Talents in University under Grant No. NCET-11-0070.

\end{document}